# Low latency carbon budget analysis reveals a large decline of the land carbon sink in 2023

Piyu Ke[1,2], Philippe Ciais[3,*], Stephen Sitch[2], Wei Li[1], Ana Bastos[4], Zhu Liu[1], Yidi Xu[3], Xiaofan Gui[5], Jiang Bian[5], Daniel S. Goll[3], Yi Xi[3], Wanjing Li[1], Michael O'Sullivan[2], Jeffeson Goncalves de Souza[2], Pierre Friedlingstein[2], Frédéric Chevallier[3]

1. Department of Earth System Science, Tsinghua University, Beijing, China

2. Faculty of Environment, Science and Economy, University of Exeter, Exeter, UK

3. Laboratoire des sciences du climat et de l'environnement CEA CNRS UVSQ U. Paris-Saclay 91191, Gif sur Yvette, France

4. Institute for Earth System Science and Remote Sensing, Leipzig University, Talstr. 35, 04103 Leipzig, Germany

5. Microsoft research

∗**Corresponding author.** E-mail: philippe.ciais@cea.fr

# Abstract

In 2023, the $CO_2$ growth rate was 3.37 ± 0.11 ppm at Mauna Loa, 86% above the previous year, and hitting a record high since observations began in 1958[1], while global fossil fuel $CO_2$ emissions only increased by 0.6 ± 0.5%[2,3]. This implies an unprecedented weakening of land and ocean sinks, and raises the question of where and why this reduction happened. Here we show a global net land $CO_2$ sink of 0.44 ± 0.21 GtC yr$^{-1}$, the weakest since 2003. We used dynamic global vegetation models, satellites fire emissions, an atmospheric inversion based on OCO-2 measurements, and emulators of ocean biogeochemical and data driven models to deliver a fast-track carbon budget in 2023. Those models ensured consistency with previous carbon budgets[2]. Regional flux anomalies from 2015-2022 are consistent between top-down and bottom-up approaches, with the largest abnormal carbon loss in the Amazon during the drought in the second half of 2023 (0.31 ± 0.19 GtC yr$^{-1}$), extreme fire emissions of 0.58 ± 0.10 GtC yr$^{-1}$ in Canada and a loss in South-East Asia (0.13± 0.12 GtC yr$^{-1}$). Since 2015, land $CO_2$ uptake north of 20°N declined by half to 1.13 ± 0.24 GtC yr$^{-1}$ in 2023. Meanwhile, the tropics recovered from the 2015-16 El Niño carbon loss, gained carbon during the La Niña years (2020-2023), then switched to a carbon loss during the 2023 El Niño (0.56 ± 0.23 GtC yr$^{-1}$). The ocean sink was stronger than normal in the equatorial eastern Pacific due to reduced upwelling from La Niña's retreat in early 2023 and the development of El Niño later[4]. Land regions exposed to extreme heat in 2023 contributed a gross carbon loss of 1.73 GtC yr$^{-1}$, indicating that record warming in 2023 had a strong negative impact on the capacity of terrestrial ecosystems to mitigate climate change.

# Introduction

The $CO_2$ growth rate in the decade 2013-2022 has averaged at 2.42 ± 0.08 ppm yr$^{-1}$. In 2023, it increased to a record high value of 3.37 ± 0.11 ppm yr$^{-1}$ at the Mauna Loa station (MLO) and 2.82 ± 0.08 ppm yr$^{-1}$ from the marine stations (MBL)[1,5] as shown in Fig. 1a. The growth rate derived from independent OCO-2 satellite observations was 3.03 ± 0.14 ppm yr$^{-1}$ (see Methods and Supplementary Fig. 1). Although during previous years, the growth rates at MLO and MBL stations have been very close (Fig. 1a), the fact that MLO was substantially higher than MBL in 2023 adds to uncertainty in understanding the carbon budget of that year. The MLO atmospheric $CO_2$ record is influenced by fluxes in Asia and the Tropics on time scales of weeks[6]. Therefore, the higher MLO growth rate could be explained by a $CO_2$ source anomaly developing in the tropics late in the year, that has not yet fully influenced other remote marine stations.

To gain insights on the carbon budget in 2023, we assessed global fossil fuel and cement emissions in 2023 from two independent sources, the Carbon Monitor project based on near real time activity data[3,7,8] and the approach from the Global Carbon Project based on preliminary energy data with partial global coverage[2]. Both estimates give a small increase of emissions of 0.1 to 1.1 % (+0.01 to +0.11 GtC yr$^{-1}$) relative to 2022 (Fig. 1b) which only explains a very small fraction of the growth rate anomaly. This implies that natural carbon sinks in the land and oceans must have been drastically reduced in 2023.

A weaker carbon sink in 2023 echoes the impact of extreme warming, globally 0.6°C above the 1991-2020 average and 1.48°C warmer than the 1850-1900 pre-industrial level[9], with extreme summer temperatures[10] and drought in the northern mid latitudes. The year 2023 was a record high for boreal forest fires in Canada, with 184961 km$^2$ of burned area, more than 2.5 times the previous recorded peak, and six times the decadal average[11]. Further, 2023 marked a transition from a long La Niña period since 2020 during which time carbon sinks are expected to be stronger than usual, towards a moderate El Niño developing after June 2023. The entire year was marked by low water storage on land observed by the GRACE satellites over most of the Northern hemisphere[12], which can cause plant water stress if soil moisture drops below a critical threshold[13]. In the tropics, the Amazon experienced an extreme drought from June to November whereas tropical Africa remained wetter than normal[14]. On the other hand, the greening of the Earth has continued in 2023 and reached peak values in mid-western USA, parts of equatorial Africa, central and south-eastern Europe, southern Brazil and northern Australia[15]. The year 2023 thus provides evidence for a decoupling between global greenness and carbon sinks over land, which deserves a regional analysis of these two variables.

To explain the record high atmospheric $CO_2$ growth rate, we developed an integrated approach using top-down and bottom-up estimates of surface $CO_2$ exchanges. Over land, we combined three dynamic global vegetation models (DGVMs)[16–22] and a high resolution atmospheric inversion assimilating OCO-2 satellite measurements[23]. The three DGVMs have been extensively validated and participated in previous global carbon budget assessments. Here, they were driven by climate reanalysis data available with a low latency in order to cover the whole year 2023 with a slightly modified protocol than in the global budget assessment (see Methods). Although we use only three DGVMs, their anomalies for previous years are close to the median of the 21 models used in previous carbon budgets assessments[2] (see Supplementary Fig. 2), which gives us confidence that our small sample of fast-track DGVM estimates can still constrain the land sink anomaly for 2023.

For the ocean carbon sink in the bottom-up budget, we used machine learning emulators of each ocean biogeochemical and data-driven model used for previous years in the global carbon budget assessment[2,25] (Supplementary Fig. 3). The emulators trained by temporal trends and patterns of the original models use as input data atmospheric $CO_2$ mixing ratio, sea surface temperature, ice cover, sea surface height, sea level pressure, sea surface salinity, mixed layer depth, wind speed and chlorophyll which are available in 2023 and allow a projection of the ocean sink in each region for the full year of 2023, since only one ocean data-driven model provided insofar low latency fluxes covering the first 8 months of the year[4].

For the top-down budget, the availability of OCO-2 observations of atmospheric carbon dioxide column concentration updated each four months allows us to cover the full year already, while most in-situ surface network measurements are not yet available. Satellite observations from OCO-2 have been shown to provide similar skill on surface $CO_2$ flux estimation than when assimilating the more accurate but sparse surface stations measurements[23]. Moreover, the OCO-2 data have better coverage than the surface network across the tropics, which is an important advantage during the year 2023 for separating $CO_2$ fluxes between the northern hemisphere and the tropics, and investigating flux anomalies between tropical continents, in particular between Amazon and South-east Asia affected by drought and Central Africa that remained wetter than normal. For the first time, the spatial resolution of our global inversion of ≈ 1° (see Methods) better matches the one of the DGVM models (0.5°), which allows us to gain more insights on regional details of $CO_2$ fluxes without the usual smoothing effect of inversions. The inversion $CO_2$ fluxes were corrected for background natural fluxes related to the river loop of the carbon cycle as in ref[26], to provide anthropogenic carbon fluxes comparable to those simulated by bottom-up models.

In the Northern Hemisphere, the occurrence of extreme forest fires in Canada caused massive emissions of $CO_2$ during the boreal late-spring and autumn. The DGVMs simulate fires from climate conditions and fuel moisture but they have strong weaknesses in capturing extreme forest fires such as those observed in Canada and simulated emissions in the range of 0.05 to 0.24 GtC $y^{-1}$ during 2023. Therefore, we used emissions based on burned area and combustion energy observed by satellites, from the Global Fire Emissions Database (GFED4.1s) and the Global Fire Assimilation System (GFAS) to assess fire emissions, and we corrected the DGVM results accordingly (see Methods). The GFED and GFAS emissions over Canada range from 0.48 to 0.68 GtC $y^{-1}$ in 2023.

We do not know if the OCO-2 inversion used for the top-down budget captures the the overall effect of fires on the carbon balance of the boreal North America region because the sampling of atmospheric $CO_2$ column by the satellite, with orbits spaced by 2000 km from each other and clouds preventing observations is to coarse to constrain emissions from single fire events. Therefore, three tests were performed in the OCO-2 inversion to estimate the sensitivity of its posterior fluxes to fire emissions: fires were either prescribed with or without vegetation regrowth, or ignored from the prior fluxes (see Methods). The tests all show very similar carbon flux anomalies in 2023 for boreal CO2 flux anomalies, showing that the inversion, even without prescribed fire emissions in the prior, can still provide a robust diagnostic of the carbon budget of this region (see Methods).

# The global carbon budget in 2023

Fig. 1 shows the bottom-up carbon budget in 2023 obtained from our estimate of fossil fuel emissions (10.20 ± 0.05 GtC $yr^{-1}$) combined with land and ocean carbon sinks from the process model emulators, and the top-down budget where the OCO-2-inversion is used to partition land and ocean sinks. The bottom-up approach does not exactly match the MBL growth rate with a difference of 0.59 ppm yr$^{-1}$ but it matches the MLO growth rate very well, within 0.04 ppm yr$^{-1}$. This translates into a budget imbalance of 1.26 GtC yr$^{-1}$ against MBL, and of 0.09 GtC yr$^{-1}$ against the MLO growth rates in 2023, comparable to the imbalance of the model ensembles used in the global carbon budget over the last years. Note that the inversion assimilating OCO-2 satellite data produces a global growth rate of 3.03 ppm yr$^{-1}$ (see Methods) which is in-between the MLO and MBL values, suggesting that MBL stations under-estimated the growth rate in 2023, as they did not yet probe the late year source anomalies in the tropics (Fig. 2).

The net land $CO_2$ flux, including land use change emissions, decreased to reach a low value of 0.44 ± 0.21 GtC yr$^{-1}$ in 2023, compared to an average of 2.04 GtC yr$^{-1}$ in the period 2010-2022, based on the bottom-up models. This is a record low value compared to previous years since 2003, both in our models (Fig. 1b) and in the models used by the global budget[2]. The OCO-2 inversion diagnosed a small sink of 0.73 ± 0.30 GtC yr$^{-1}$ for the starting El Nino similar to the previous El Niño of 2015-2016 which was nevertheless more extreme than the moderate El Niño starting in June 2023. We will need to acquire fluxes until early 2024 to cover the entire period of the current El Niño and compare it to the 2015-2016 event.

In 2023, the ocean carbon sink increased by 0.10 GtC yr$^{-1}$ compared to the year 2022 in our bottom-up approach, to reach a value of 2.60 ± 0.72 GtC yr$^{-1}$. The inversion gives an ocean sink of 2.33 ± 0.20 GtC yr$^{-1}$, in close agreement with our bottom-up models emulators. This increase in the ocean carbon sink was mainly due to the fading La Niña and the developing El Niño, which decreased CO2 sources in the tropical Pacific, while high SSTs reduced the sink in the Northeastern Atlantic[4].

# Regional anomalies

To gain insights on which regions caused the large drop of the land sink and a coincident increase of the ocean sink in 2023, we analyzed the spatial patterns of flux anomalies for each quarter of the year 2023 in the bottom-up models and in the OCO-2 inversion using as a reference period 2015-2022 which corresponds to the period covered by the inversion. The results are displayed in Figure 2 and quarterly fluxes in 2023 within the distribution of previous years is shown in Supplementary Fig. 4. See Supplementary Figure S6 for the distribution of the RECCAP2 regions.

Over the ocean, the most notable increases in carbon uptake were observed in the Pacific Ocean and parts of the Southern Ocean, consistent between the ocean models emulators and the inversion (Fig. 2). Particularly, the increased uptake was most pronounced in the eastern equatorial Pacific, consistent with suppressed upwelling of carbon rich waters during the developing El Niño[27]. The carbon sinks in the Arctic Ocean, the Indian Ocean and coastal oceans remained relatively unchanged. There is a divergence in the Southern Ocean, where ocean model emulators suggest a slight decrease, but the OCO-2 inversion indicates an increase (Supplementary Fig. 5, Supplementary Table 2). On a quarterly basis, the growth of the global ocean sink was mainly observed during the two last quarters of 2023, with the arrival of El Niño conditions.

Regional quarterly land flux anomalies in Fig. 2 indicate large spatial contrasts, which are roughly consistent between the inversion and the bottom-up models ($R^2$ = 0.33 for land, $R^2$ =0.30 for ocean), In general, the inversion shows dipoles of land sources and sinks of larger magnitude than the bottom-up models. The northern land uptake which is usually peaks in summer (JAS) (1.08 GtC month$^{-1}$ in the inversion, 0.52 GtC month$^{-1}$ in the DGVMs north of 20°N during 2015-2022) was lower in 2023 with with abnormal sources emerging in Central Europe, western Russia, central America (Fig. 2 and Supplementary Fig. 4 and Supplementary Fig. 7). Yet, the JAS period in 2023 shows a large sink in the north western part of North America of 0.11 GtC month$^{-1}$ in the inversion, but not in the DGVMs (0.01 GtC month$^{-1}$), over the region from 180° to 100°W and 40°N to 70°N). The prevalence of a net sink despite extreme fire emissions in Canada suggests that the fire emissions were offset by a region-wide exceptional summer uptake, as corroborated by maximum greening anomalies[15]. As stated above, the OCO-2 inversion JAS sink anomaly in North America was very similar between three inversions with different assumptions about prior fire emissions, suggesting that the OCO-2 data can robustly assess the flux anomalies at this spatial scale (Supplementary Table 3)

In the fourth quarter of 2023 (OND), the land carbon sink was much lower than the average of previous years. During this quarter, large abnormal carbon losses appeared in the Amazon (-0.14 GtC month$^{-1}$ in the inversion, -0.33 GtC month$^{-1}$ in the DGVMs) and were the largest contributor to the drop of the annual global land sink in 2023. In Africa, the inversion shows an abnormal loss of 0.17 GtC month$^{-1}$ and the DGVMs a loss of 0.07 GtC month$^{-1}$ during OND. Abnormal uptake in central and eastern Africa that experienced wetter conditions was offset by sources in southern Africa[12]. In Southeast Asia, both bottom-up and top-down show a net sink approaching zero in OND, ranging from 0.003 GtC month$^{-1}$ to 0.01 GtC month$^{-1}$ (Supplementary Fig. 4).

Overall, it is the compounding coincidence of a large abnormal source in the Tropics contributed by the Amazon drought offsetting higher sinks in central and eastern Africa and Western North America, and a weak summer uptake in the rest of the northern Hemisphere that explains the cancellation of the global land sink in 2023 (Fig. 3, Supplementary Table 1). Interestingly Liu et al.[28] noted that in 2021, despite strong La Niña conditions, the land sink was not as large as during previous strong La Niña events and explained this phenomenon by a smaller northern hemisphere uptake, being compensated by a higher tropical sink. Here in 2023, we see a continuation and an amplification of a weakening northern sink, coinciding with a large abnormal carbon loss in the Tropics with the moderate El Niño arriving in the second half of the year. Since 2015, the northern hemisphere land sink north of 20°N continuously declined by a factor of two to reach 1.13 ± 0.24 GtC yr$^{-1}$ in 2023 (mean of bottom-up and top-down) while the tropics recovered from an extreme carbon loss after the 2015-16 El Niño and remained a sink during the long sequence of wet and cool La Niña years from 2020 to early 2023, then switched to a carbon source during the moderate El Niño in the end of 2023 (Fig. 3).

Intriguingly, 2023 shows a strong negative relationship between a reduced global land sink and a record-high global greening level[15]. Yet, when looking at the regional scale, we found the expected positive spatial relationships between greening and carbon uptake (Supplementary Fig. 8). In North America, local burned forest areas where browning was observed have however lost a disproportionate amount of carbon compared to the diffuse gains over the large area of non-burned forests with widespread greening. Therefore, it is the fact that carbon losses are very intense over very small disturbed areas where browning is observed that leads to an apparent global decoupling between greening and carbon sinks, but locally carbon losses still associate with browning and gains to greening. A second reason for the local decoupling between greening and carbon sinks is that burned forests recover to a higher grenness very shortly after fires[29], while they recover carbon at a very slow rate[30].

We also found in 2023 that many areas with weaker carbon sinks or higher carbon sources systematically associate with water storage deficits observed by the GRACE satellites (Fig. 3) in water limited regions. Some regions where the vegetation is not limited by water availability such as the Pacific Northwest yet experienced a water deficit and higher carbon sinks in 2023. Thus, care should be taken about systematically expecting that low water availability turns ecosystems into carbon sources - this is only true over regions that lie close to their water limitation point[13].

Given the occurrence of record breaking temperatures in 2023, we finally analyzed the carbon flux anomalies over all grid cells, grouped per percentile intervals of temperature anomalies from the reference period 1991-2020. The results shown in Fig. 3 indicate that extremely hot grid cells (above the 95$^{th}$ percentile) altogether contributed a gross carbon loss of -1.73 GtC yr$^{-1}$ over the globe, including -0.27 GtC yr$^{-1}$ in the northern mid latitudes (20-60° N) and -1.36 GtC yr$^{-1}$ over the Tropics (20°S - 20°N). The 95$^{th}$ hottest grid cells account for 29.57% of the global gross carbon loss and have no gross carbon sink in 2023, although they cover only 8.61% of the global land area, showing the disproportionate impact of extremely hot temperatures on negating carbon uptake by the land vegetation.

This result is alarming as temperatures continue to keep very high values in 2024. It is too early to conclude on a durable collapse of the land sink in the aftermath of 2023. Yet, forests burned in Canada will not completely restore their carbon stocks for the next decades, given that it takes about 100 years for boral trees to recover their initial biomass[31]. Forests in the wet tropics have proven vulnerable to previous extreme droughts, but they were also found to recover quickly, e.g., within a few years in most regions after the severe past El Niño of 2015-16[32]. However, forest resilience has been decreasing over time in the Amazon[33]. It is unknown whether a new regime of hotter droughts in the Tropics will induce a shift in tree mortality that could turn these critical carbon rich systems into long term carbon sources. As per northern forests, it seems that recurrent hot conditions had already begun to weaken their carbon uptake at least since 2021[28]. The resilience of these systems to droughts and droughts-related impacts (insect attacks) combined with future management practices such as harvest rates will determine the near-future stability of the northern sink.

The observation that 2023 had an exceptionally weak land sink despite being only a moderate El Niño constitutes a test bed for Earth System models which lack processes causing rapid carbon losses, such as extreme fires and climate-induced tree mortality in their projections, and may thus be too optimistic for estimating remaining carbon budgets[34]. If very high warming rates continue in the next decade and negatively impact the land sink as they did in 2023, it calls for urgent action to enhance carbon sequestration and reduce greenhouse gasses emissions to net zero before reaching a dangerous level of warming at which natural $CO_2$ sinks may no longer provide to humanity the mitigation service they have offered so far by absorbing half of human induced $CO_2$ emissions.

# Methods

**Atmospheric $CO_2$ growth rate (CGR).** We used the monthly time series of globally averaged marine surface (MBL) atmospheric $CO_2$ concentration covering the period from January 1979 to February 2024, and the MLO station data from March 1958 to April 2024, both provided by NOAA's Global Monitoring Laboratory (NOAA/GML)[1,5]. The annual MBL growth rate is determined by averaging the most recent December and January months, corrected for the average seasonal cycle, and subtracting the average of the same period in the previous year

([https://gml.noaa.gov/ccgg/trends/gl_gr.html](https://gml.noaa.gov/ccgg/trends/gl_gr.html)). For the MLO data, the annual mean growth rate is estimated using the average of the most recent November-February months, corrected for the average seasonal cycle, and subtracting the same four-month average of the previous year centered on January 1st (https://gml.noaa.gov/ccgg/trends/gr.html).

**Global fire $CO_2$ emissions.** Both the Global Fire Emissions Database version 4.1 including small fire burned area (GFED4.1s)[35] and the Global Fire Assimilation System (GFAS, https://atmosphere.copernicus.eu/global-fire-monitoring) from Copernicus Atmosphere Monitoring Service (CAMS) are used to derive monthly global fire $CO_2$ emissions. GFED4.1s combines satellite information on fire activity and vegetation productivity to estimate the gridded monthly burned area and fire emissions, and has a spatial resolution of $0.25° \times 0.25°$[35]. Note that GFED4.1s fire emissions in 2017 and 2023 are from the beta version. The GFAS assimilates fire radiative power (FRP) observations from satellites to produce daily estimates of biomass burning emissions. We aggregated the GFAS daily fire emissions into monthly emissions.

**Terrestrial $CO_2$ fluxes.** Terrestrial carbon fluxes are derived from the mean of three Dynamic Global Vegetation Models (DGVMs), specifically ORCHIDEE, JULES and OCN. The methodology for estimating terrestrial carbon fluxes for the period 2010-2023 aligns with the TRENDY protocol, used in Global Carbon Budgets[24], albeit with modifications due to the use of ERA5 climate forcing, and the fact that land-use forcing is not updated with such short latency. ERA5 forcing has been available since 1940[36], but preliminary simulations with DGVMs showed some issues with precipitation forcing before 1960. The simulations performed here correspond to the S2 experiment, starting from steady-state in 1960, with time varying climate and $CO_2$ forcing, and land-use fixed at 2010, as described in ref.[37]

Since the model simulations start from a steady-state in the 1960s, they cannot account for the carbon balance before the industrial revolution, which leads to an underestimation of the mean trend compared to standard DGVMs. Therefore, we calibrated them using the 2019-2022 TRENDY DGVMs from the Global Carbon Budget 2023[2]. This was done by calculating for each grid cell the median of the carbon flux from the TRENDY models during 2015-2022 and the mean of each of the DGVMs used in this study, then subtracting from each DGVM on each grid the difference so that they match the median flux of the TRENDY models. In other words, our DGVMs are used for predicting interannual anomalies of CO2 fluxes.

**Ocean $CO_2$ fluxes.** Ocean carbon fluxes are derived from a suite of emulators based on both biogeochemical models and data-driven models. We updated estimates from 5 Global Ocean Biogeochemical Models and 8 data products included in the Global Carbon Budget 2022 to create a near-real-time framework[25]. This update employs Convolutional Neural Networks (CNNs) and semi-supervised learning techniques to capture the non-linear relationships between model or product estimates and observed predictors. As a result, we obtain a near-real-time, monthly grid-based dataset of global surface ocean fugacity of $CO_2$ and ocean-atmosphere $CO_2$ flux data, extending to December 2023. More details are given in ref[25] and in Supplementary.

**Anthropogenic $CO_2$ emissions.** For the period from 2010 to 2022, we used global fossil fuel and cement $CO_2$ emissions estimates from the latest edition of the global carbon budget[2]. For 2023, we used the average of emissions from the Carbon Monitor project based on near-real-time activity data from 6 sectors[3,7,8] and from an updated estimate using the same methodology that the global Carbon Budget[2], i.e. based on energy data available by the time of the publication and projections for countries with no available data. The Carbon Monitor data give a global emission of +0.1% compared to 2022 while the Global Carbon Budget approach gives a global emission of 1.1% compared to 2022.

**Atmospheric inversion.** We used a global high resolution atmospheric inversion driven by the OCO-2 satellite atmospheric $CO_2$ concentration data[38] called CAMS. It can deliver global estimates of weekly greenhouse gas fluxes with a typical 4-month latency, now at a resolution of 0.7 degree in latitude and 1.4 degree in longitude. The product used here is an intermediate version between version FT23r3 and the coming FT23r4. It follows the usual production and quality control process of the CAMS product. It covers the OCO-2 period from 2015 to December 2023, and its mean fluxes and anomalies are close to the median of 14 inversions used in previous assessments[2] (see Supplementary Fig. 9).

The underlying transport model was nudged towards horizontal winds from the ERA5 reanalysis. The inferred fluxes were estimated in each horizontal grid point of the transport model with a temporal resolution of 8 days, separately for day-time and night-time. The prior values of the fluxes combine estimates of (i) gridded monthly fossil fuel and cement emissions (GCP-GridFED version 2023.1[39]) extended to year 2023 following Chevallier et al. (2020)[40] using the emission changes reported by https://carbonmonitor.org/, together with anomalies in retrievals of $NO_2$ columns from the Tropospheric Monitoring Instrument (TROPOMI, offline and processing and RPRO when available, van Geffen et al., 2019[41]), (ii) monthly ocean fluxes (Chau et al. 2023[4], 2024[42]), 3-hourly (when available) or monthly biomass burning emissions (GFED 4.1s) described in van der Werf et al. (2017)[35] and climatological 3-hourly biosphere-atmosphere fluxes taken as the 1981-2020 mean of a simulation of the ORganizing Carbon and Hydrology In Dynamic EcosystEms model, version 2.2, revision 7262 (ORCHIDEE, Krinner et al. 2005[21]).  The variational inversion accounts for spatial and temporal correlations of the prior errors, resulting in a total 1-sigma uncertainty for the prior fluxes over a full year of 3.0 GtC·yr$^{-1}$ for the land pixels and of 0.2 GtC·yr$^{-1}$ for the marine pixels.

**Land fluxes per percentile of land temperature anomalies from the reference period 1991-2020.** We utilized 2m temperature data spanning 1960-2023 from monthly ERA5 data[36]. Monthly gridded data from 1991-2020 were sorted for each grid cell by temperature, categorized into 20 percentiles. Subsequently, temperature data from 2010-2022 and 2023 were used to identify grid cells falling within these percentiles, calculating their respective areas. Concurrently, matching grid cells were identified from three NRT DGVMs employed in this study to compute land fluxes.

**Ranked greening and net land carbon fluxes.** We use Moderate Resolution Imaging Spectroradiometer (MODIS) NDVI data[43] to assess vegetation greenness in 2023. We rank the NDVI data for each grid from 2010 to 2023, obtaining the position of each grid's 2023 NDVI within this period. For land carbon fluxes, we utilize data from 3 NRT DGVMs and OCO-2 inversion. We rank the 2023 flux for each grid within the periods 2010-2023 for the DGVMs and 2015-2023 for the OCO-2 inversion.

# Data availability

The data from Global Carbon Budget 2023 are available at https://www.icos-cp.eu/science-and-impact/global-carbon-budget/2023. The Carbon monitor fossil fuel emissions dataset is available at https://carbonmonitor.org/. The GFED 4.1s fire emissions dataset is available at geo.vu.nl/~gwerf/GFED/GFED4/. The GFAS fire emissions dataset is available at

https://atmosphere.copernicus.eu/global-fire-monitoring/. The ERA5 monthly averaged data is available at

https://cds.climate.copernicus.eu/cdsapp#!/dataset/reanalysis-era5-single-levels-monthly-means?tab=overview. The Multivariate ENSO index is available at https://www.psl.noaa.gov/enso/mei. The MODIS NDVI data is available at https://lpdaac.usgs.gov/products/mod13c2v006/.

# Author contributions

P.C. and P.K. designed the research; P.K. and P.C. performed the analysis; P.K., P.C., S.S., A.B., X.G., M.O. and F.C. collected and analyzed the research data; P.K. and P.C. created the first draft of the paper; all authors contributed to the interpretation of the results and to the text.

**Conflict of interest statement.** None declared.

# Acknowledgements

P.C., F.C., A.B. S.S., P.F. acknowledge support from the European Space Agency Climate Change Initiative RECCAP2 project 1190 (ESA ESRIN/ 4000123002/ 18/I-NB) and ESA Carbon-RO (4000140982/23/I-EF). A.B. thanks Evgenii Churiulin for support in setting up the OCN model simulations.

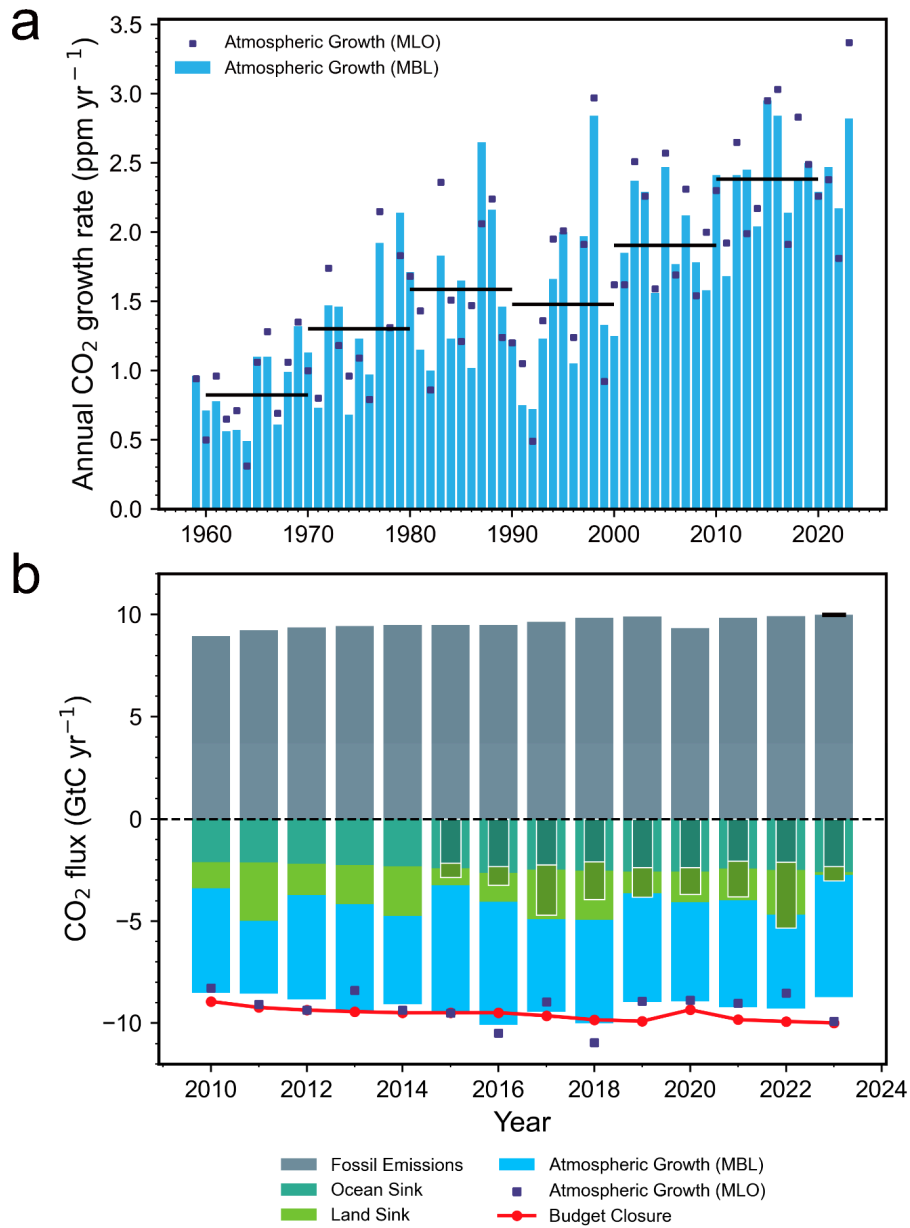

**Fig. 1 Atmospheric $CO_2$ growth rate from 1960-2023 and carbon budget from 2010-2023.** (a) Growth rate from marine boundary layer surface stations (MBL, blue bars) and the Mauna Loa station (MLO, dark blue squares). (b) Global $CO_2$ budget obtained with historical fossil fuel and cement $CO_2$ emissions and our estimates of land and ocean sinks in 2023, and the MBL / MLO $CO_2$ annual growth rates. Our estimates are based on simulations by emulators of the ocean sink, simulations of the land sink by three dynamic vegetation models forced by low latency climate input data ( their mean sink in 2019-2022 being adjusted to equal the median of 16 models used in the latest Global Carbon Budget edition). The ocean and land sinks in the inside bars are from the OCO-2 high resolution atmospheric inversion. The difference between the stacked bars at the bottom and the red curve (-1 x Fossil emissions) is the imbalance of the budget.

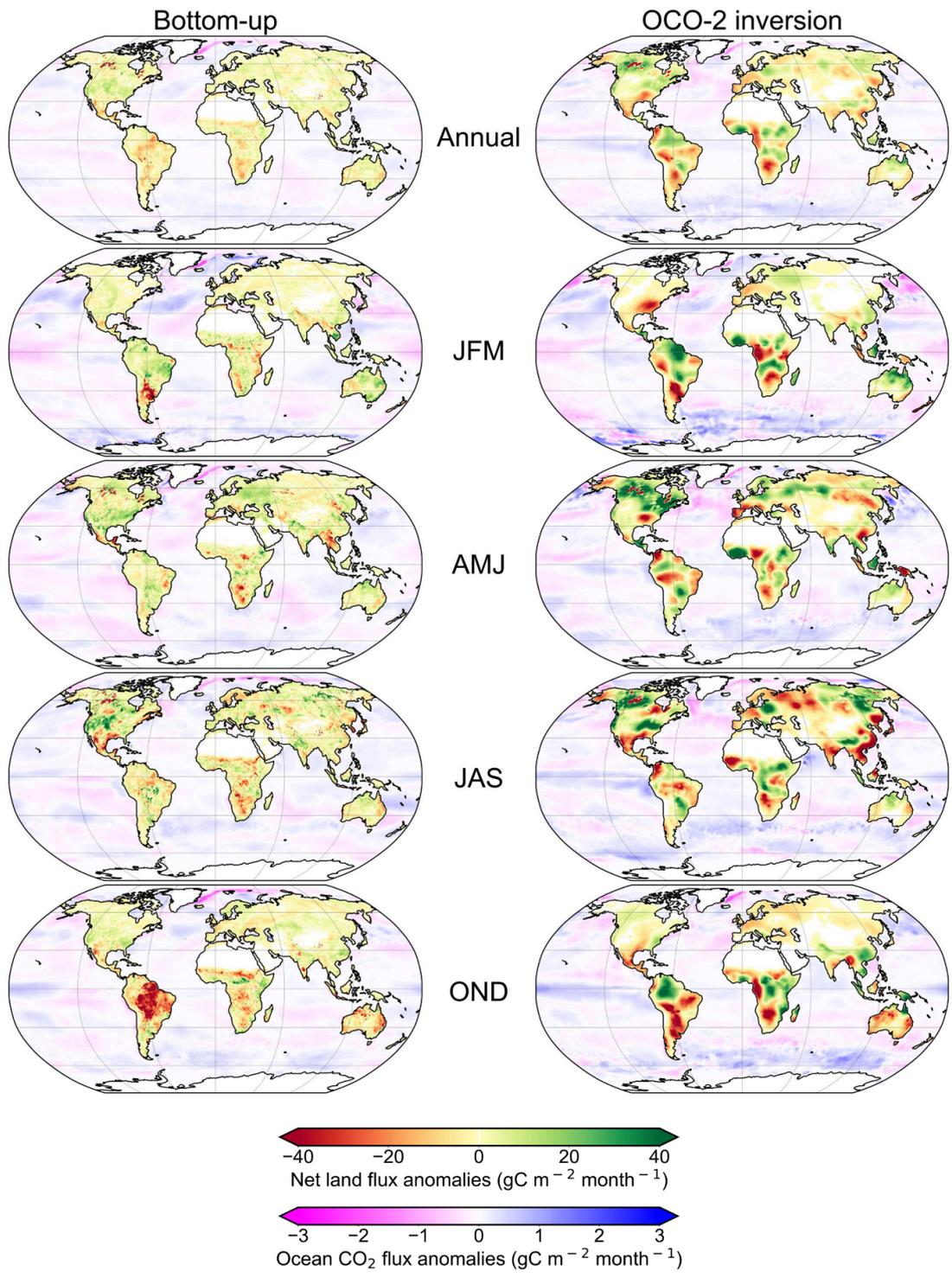

**Fig. 2 Net land and ocean CO₂ flux anomalies for each quarter in 2023 compared with the 2015-2022 average for bottom-up models (left column) and the OCO-2 inversion (right column).**

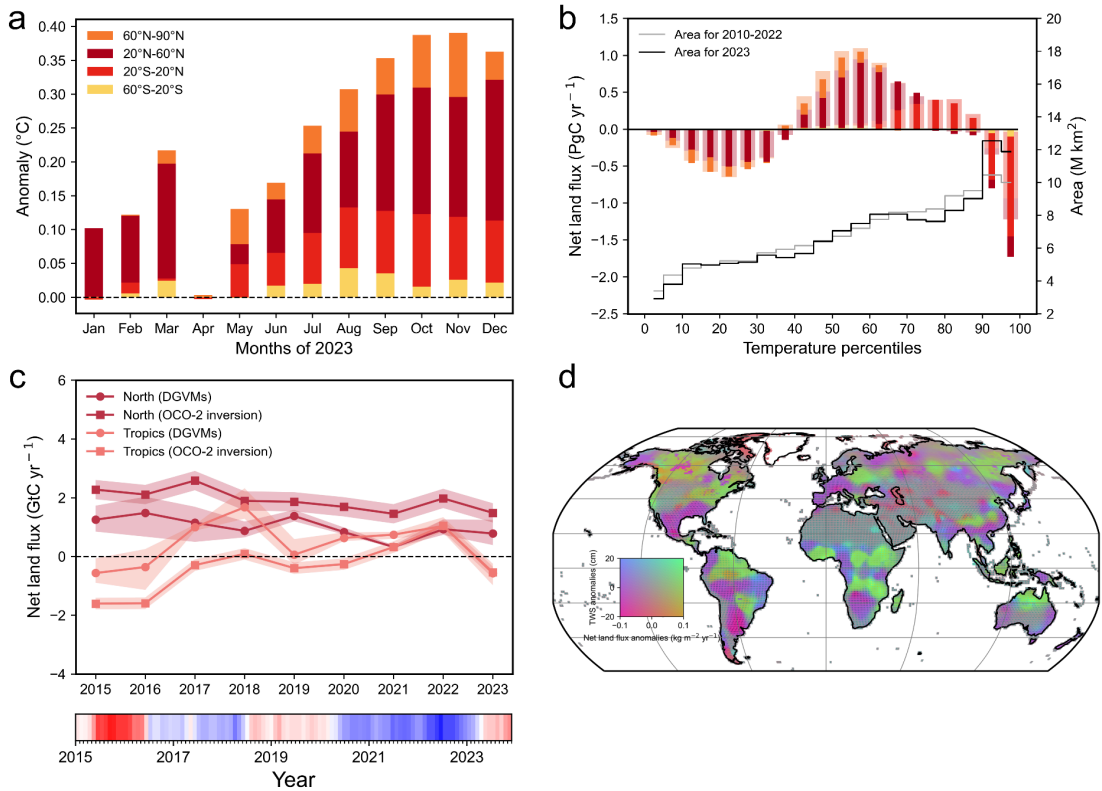

Fig. 3 (a) Land temperature warming anomaly each month in 2023 from the reference period 1991-2020 in four large latitude bands. (b) Land $CO_2$ fluxes in 2010-2022 (wide bars) and in 2023 (inside bars) for each percentile of land temperature calculated during the reference period 1991-2020. The colors are for the same latitude bands than a. The land area in each percentile is represented by the gray line in 2010-2022 and by the black line in 2023. (c) Changes in the declining northern land sink and the variations of tropical land flux with sources in 2015-16 and 2023, from the bottom-up DGVMs and the OCO-2 inversion, during the period from 2015 to 2023. The ENSO index from NOAA Physical Sciences Laboratory (https://www.psl.noaa.gov/enso/mei) is represented in the bottom, with the extreme El Niño of 2015-16, the La Liña from mid 2020 to mid 2023, and the moderate El Niño of the second half of 2023. (d) Bivariate maps of GRACE TWS anomalies and net land flux anomalies from OCO-2 inversion in 2023 compared with the 2015-2022 average.

Supplemental Information for

# Low latency carbon budget analysis reveals a large decline of the land carbon sink in 2023

Piyu Ke[1,2], Philippe Ciais[3,*], Stephen Sitch[2], Wei Li[1], Ana Bastos[4], Zhu Liu[1], Yidi Xu[3], Xiaofan Gui[5], Jiang Bian[5], Daniel S. Goll[3], Yi Xi[3], Wanjing Li[1], Michael O'Sullivan[2], Jeffeson Goncalves de Souza[2], Pierre Friedlingstein[2], Frédéric Chevallier[3]

1. Department of Earth System Science, Tsinghua University, Beijing, China
2. Faculty of Environment, Science and Economy, University of Exeter, Exeter, UK
3. Laboratoire des sciences du climat et de l'environnement CEA CNRS UVSQ U. Paris-Saclay 91191, Gif sur Yvette, France
4. Institute for Earth System Science and Remote Sensing, Leipzig University, Talstr. 35, 04103 Leipzig, Germany
5. Microsoft research

∗**Corresponding author.** E-mail: philippe.ciais@cea.fr

**This PDF file includes:**
    Supplementary Texts
    Supplementary Figures 1 to 9
    Supplementary Tables 1 to 3

# Supplementary Texts

1. **Global annual growth rates of atmospheric $CO_2$ from different atmospheric observations and bottom-up models**

We compared the atmospheric $CO_2$ growth rate from 2010 to 2023 based on in-situ observations from 40 marine boundary layer background stations, calculated by NOAA ESRL[1,2] as the year on year difference between smoothed observations between November-February averaged across all the stations (blue bars), from the Mauna Loa station (purple squares), from the assimilation of global OCO-2 satellite observations of column $CO_2$ concentration measurements, about 300,000 10-second-averaged retrievals each year, by the inversion used in this study (brown dots), and by the bottom-up approach, that is, not using atmospheric measurements and the growth rate is predicted from the difference between fossil fuel $CO_2$ emissions minus the land sink from three DGVM models, minus the ocean sink from ocean model emulators (red dots) (Supplementary Figure 1).

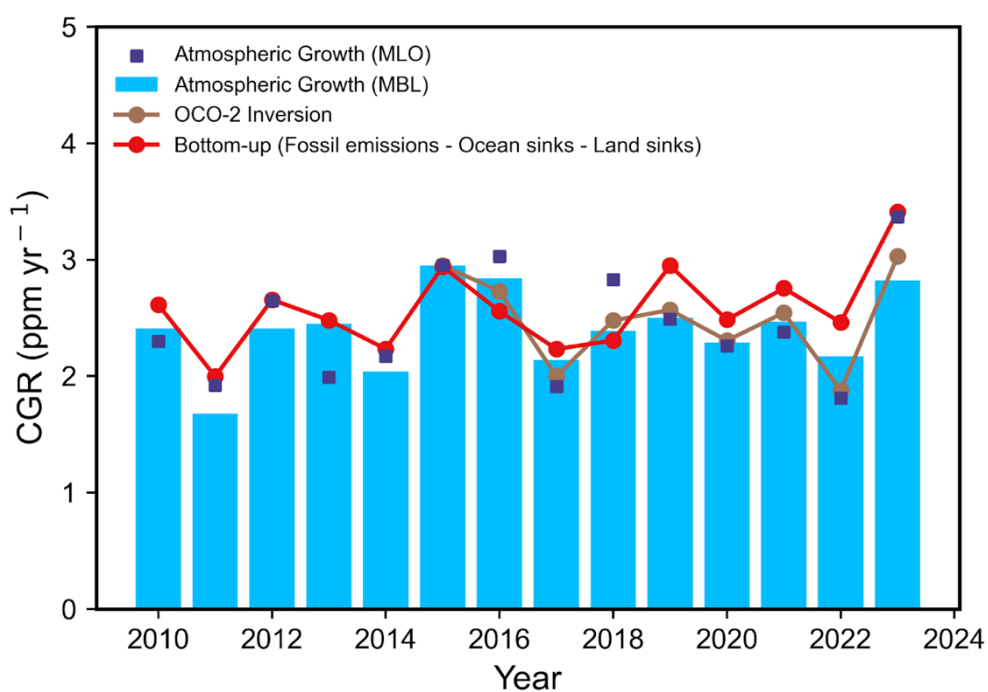

**Supplementary Figure 1 $CO_2$ growth rate from 2010 to 2023 from different approaches.** Growth rates are derived from marine boundary layer surface stations (MBL, blue bars), the Mauna Loa station (MLO, dark blue squares), assimilation of global OCO-2 satellite observations of column $CO_2$ concentration measurements (brown dots), and a bottom-up approach where atmospheric measurements are excluded. The bottom-up approach predicts growth rates from the difference between fossil fuel $CO_2$ emissions and the sum of the land sink from three DGVM models and the ocean sink from ocean model emulators (red dots).

## 2. Maps of monthly air-sea $CO_2$ fluxes from emulators of biogeochemical and data driven ocean models

We utilized a deep learning technique for near-real-time estimation of oceanic carbon monthly gridded fluxes. This method integrates year, month, latitude, longitude, and nine environmental factors as predictors, targeting predictions for each GOBM model or ocean data product. We use monthly data from 5 GOBMs and 8 data products from the Global Carbon Budget 2022[3], covering data up to the end of 2021. The $fCO_2$ output from each GOBM or data product is provided at a 1° × 1° monthly resolution. Our predictive variables include a range of biological, chemical, and physical factors typically linked to fluctuations in $fCO_2$. These variables are sea surface temperature (SST), sea ice fraction (ICE), sea surface salinity (SSS), atmospheric $CO_2$ mole fraction ($xCO_2$), mixed-layer depth (MLD), sea surface height (SSH), chlorophyll a (chl a), sea level pressure (SLP), and wind speed. All data are bilinearly interpolated to a 1° × 1° monthly resolution to align with our $fCO_2$ targets and updated to Dec. 2023. Since the $xCO_2$ data is only available up to the end of 2022, and to meet the requirement for near-real-time data, we gather global average marine boundary layer surface monthly mean atmospheric $CO_2$ data updated to Dec. 2023. We use a light gradient boosting machine (LightGBM)[4] model to establish a relationship between the year, month, latitude, longitude, mean atmospheric $CO_2$ data, and $xCO_2$. We used data from 1979-2021, divided into training and validation datasets in an 8:2 ratio. Early stopping was implemented with LightGBM, and testing on 2022 data yielded a test RMSE of 1.74, reflecting roughly a 0.5% prediction error. This approach allows us to extend the $xCO_2$ data to near-real-time.

The data are formatted into a 180x360 grid and subdivided into 18x18 patches for computational efficiency. We trained the model on labeled data points, calculating the Root Mean Square Error (RMSE) between labels and predictions as the supervised loss (LsL_sLs). To ensure prediction stability on unlabeled data, we used pseudo-labeling by predicting with 10% of features removed as pseudo-labels, then predicting again with 30% of features removed, calculating the RMSE as the unsupervised loss (LuL_uLu). The model was updated using the weighted sum of LsL_sLs and LuL_uLu through backward propagation. Our model architecture combines a multi-layered Convolutional Neural Network (CNN)[5] and linear models[6], designed to process the input data efficiently. The input layer has a dimension of 18x18x13, maintained across all CNN and linear layers. The CNN hidden layers have dimensions of 13, 64, and 64, learning spatial hierarchies, while the linear layers have dimensions of 64, 64, and 1, performing linear transformations for predictions. The output layer has a dimension of 1, representing the predicted oceanic carbon $fCO_2$ value. CNNs, inspired by human visual perception, consist of Convolutional Layers, Rectified Linear Unit (ReLU) Layers, and Fully Connected Layers, which work together to process and transform input data into predictions. This architecture ensures accurate and efficient prediction of oceanic carbon fluxes.

## 3. Low latency terrestrial CO$_2$ fluxes from land surface models and ocean CO$_2$ fluxes from emulators

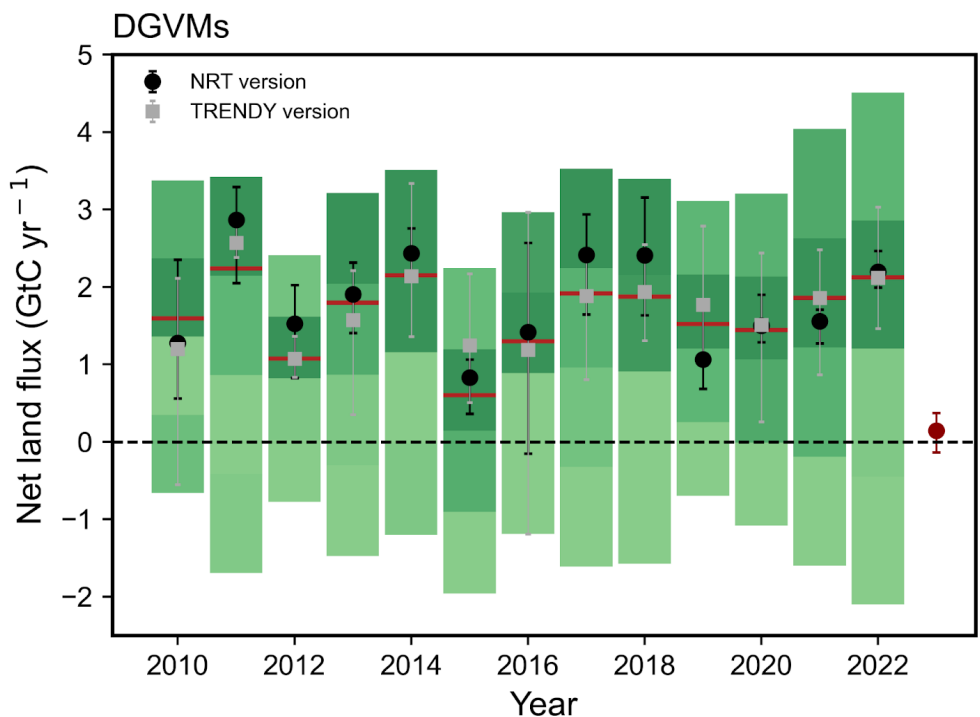

**Supplementary Figure 2 Comparison of global net land CO$_2$ fluxes between the 3 DGVMs used in this study (black dots for NRT version, a red dot for NRT version in 2023, gray squares for TRENDY version) and the distribution of 21 TRENDY models (green bars, with color intensity representing the number of DGVMs falling into each interval) used in latest Global Carbon Budget edition**[7]**.** Note that the 3 DGVMs have been biased corrected to have the same net land sink than the average of TRENDY models during 2019-2022, but their anomalies are feely calculated for all years including in 2023.

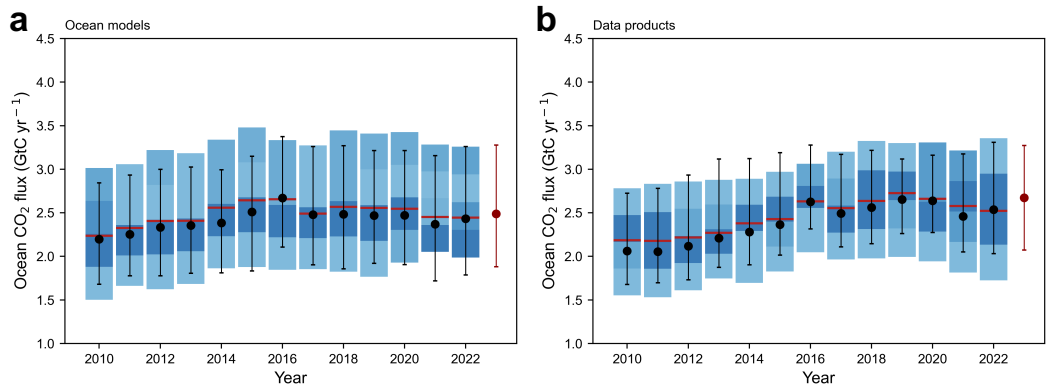

**Supplementary Figure 3** Comparison of global ocean CO₂ fluxes between the emulators of ocean models used in this study (black dots for 2010-2022, red dots for 2023) and the distribution of 5 mechanistic ocean biogeochemistry models (a) and 8 data driven models (b) (blue bars, with color intensity representing the number of models falling into each interval) used in latest Global Carbon Budget edition[7].

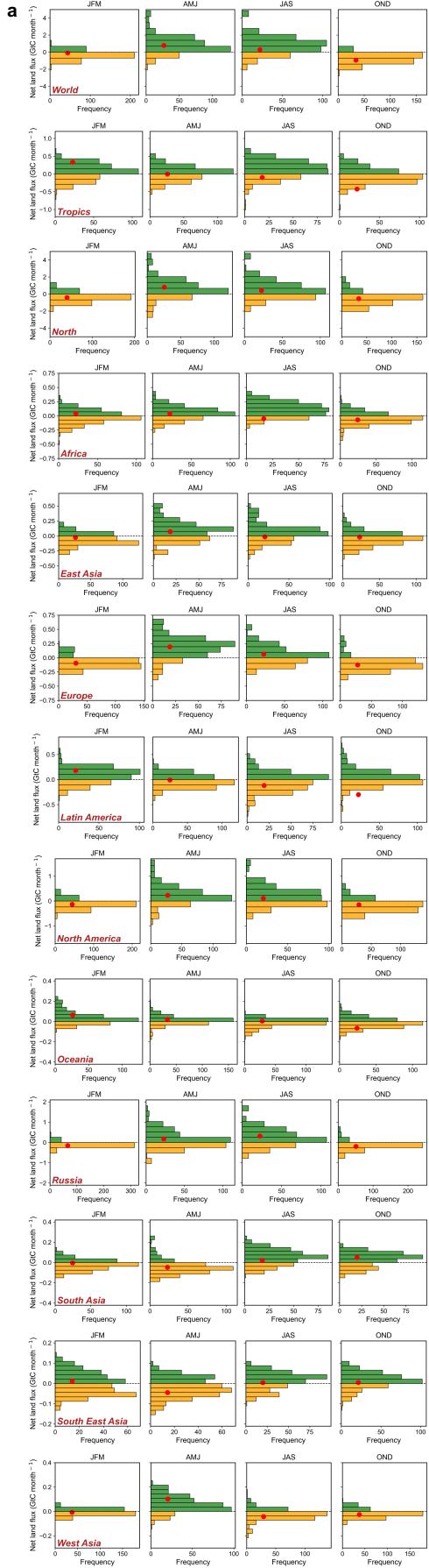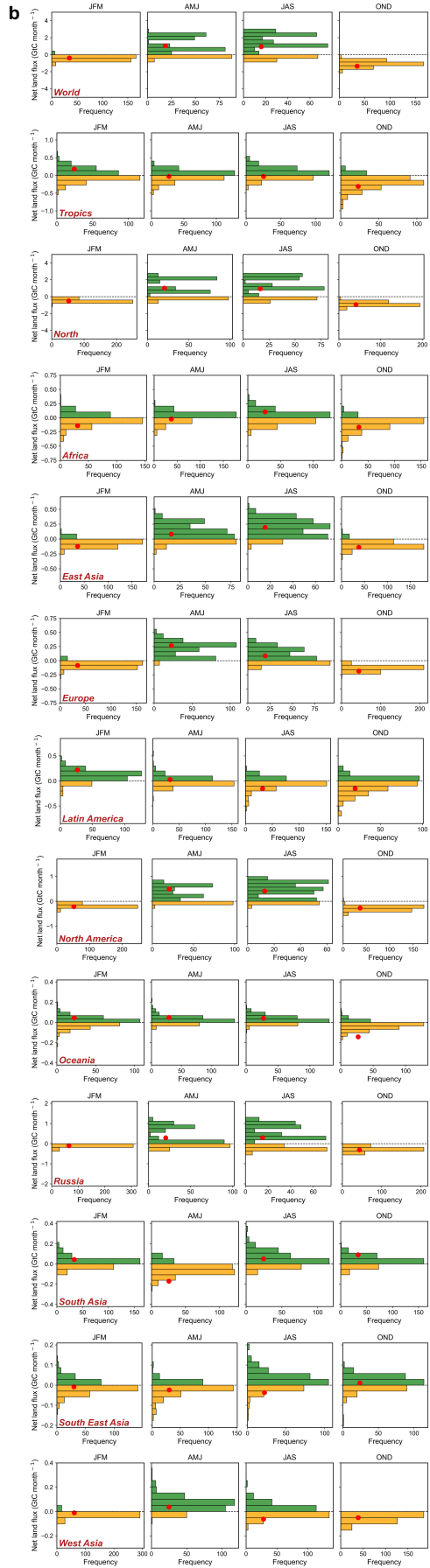

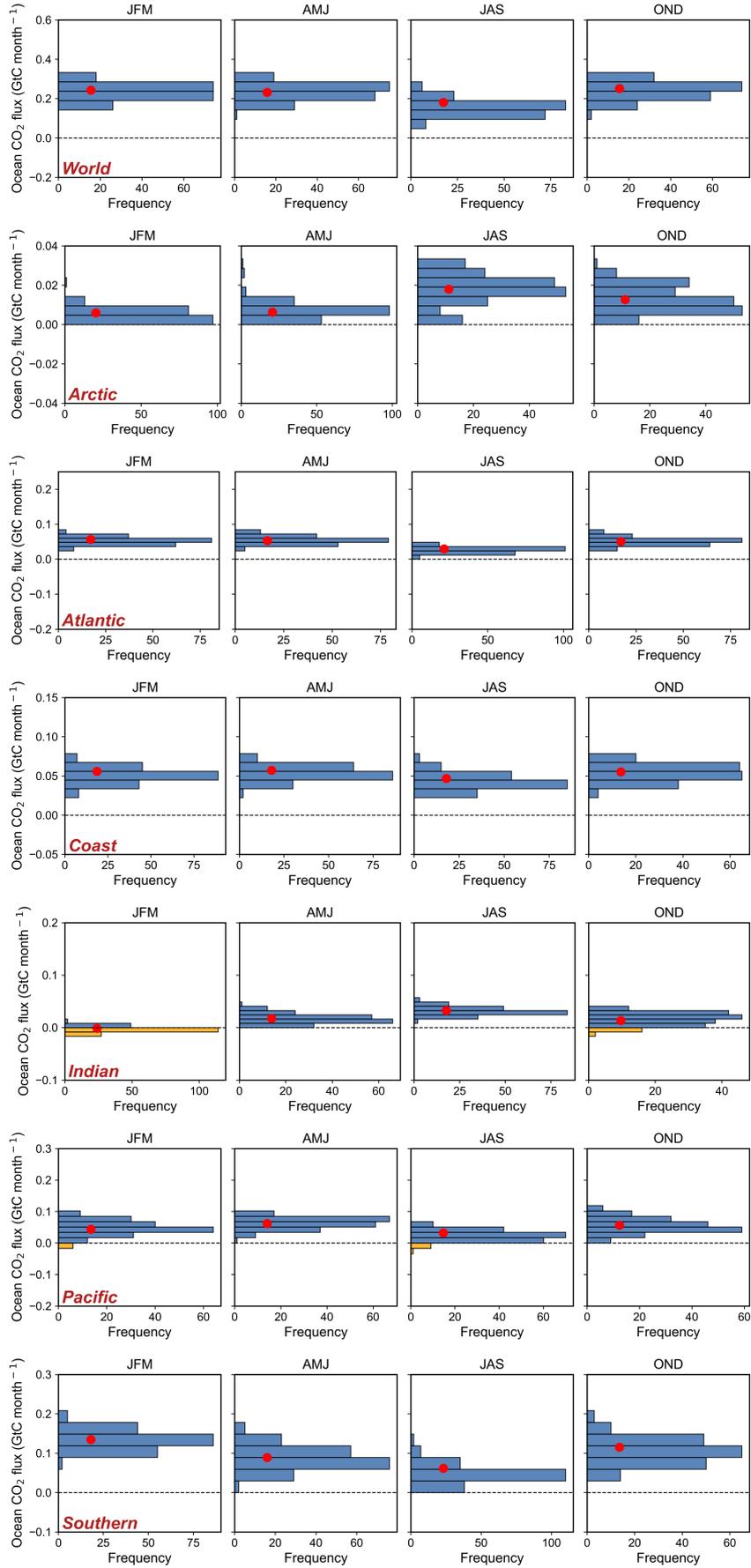

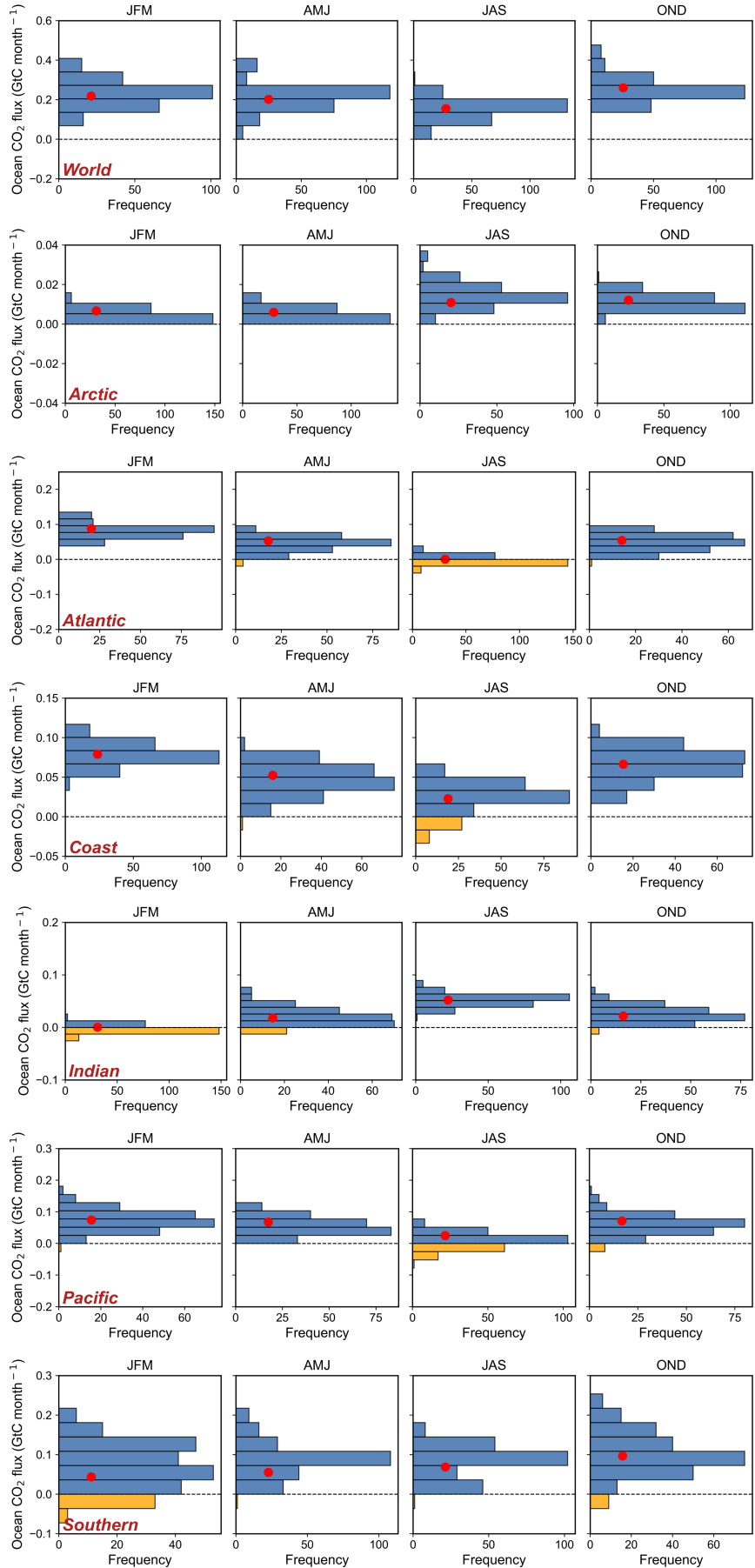

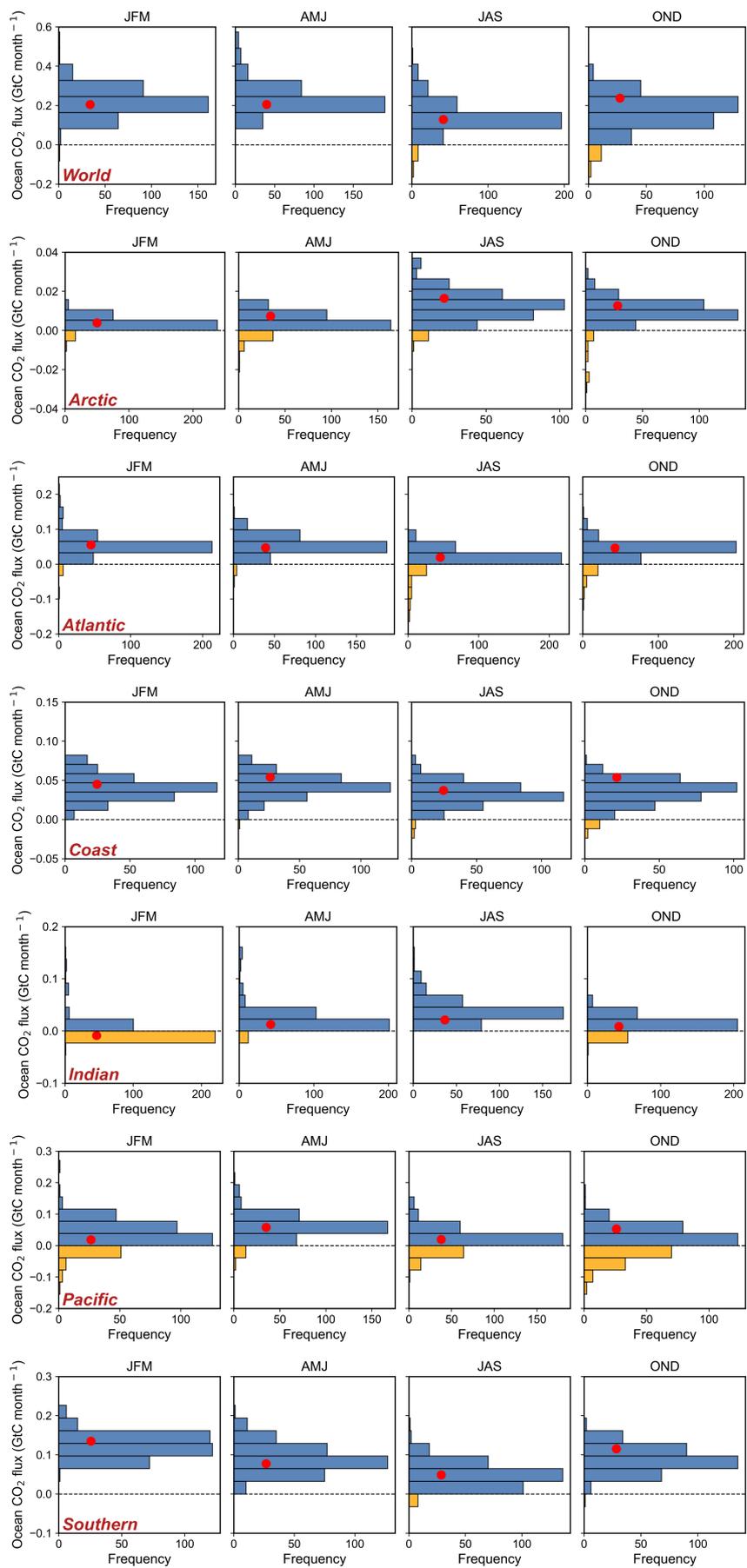

**Supplementary Figure 4 Quarterly land and ocean fluxes for the RECCAP2 regions in 2023 and the distributions during 2015-2022.** (a) Quarterly land flux in 2023 from 3 NRT DGVMs (red dots) for each RECCAP2 land region in the distribution of the flux from the DGVM models. (b) Quarterly land flux in 2023 from the OCO-2 inversion (red dots) for each RECCAP2 land region in the distribution of the flux from the inversion models. (c) Quarterly ocean flux in 2023 from the ocean data products emulators (red dots) for each RECCAP2 ocean region in the distribution of the flux from the data products models. (d) Quarterly ocean flux in 2023 from the ocean GOBMs emulators (red dots) for each RECCAP2 ocean region in the distribution of the flux from the GOBMs. (e) Quarterly ocean flux in 2023 from the OCO-2 inversion for each RECCAP2 ocean region in the distribution of the flux from the inversion models. Distributions are calculated during all previous years in the period 2015-2022. See Supplementary Figure S6 for the distribution of the RECCAP2 regions.

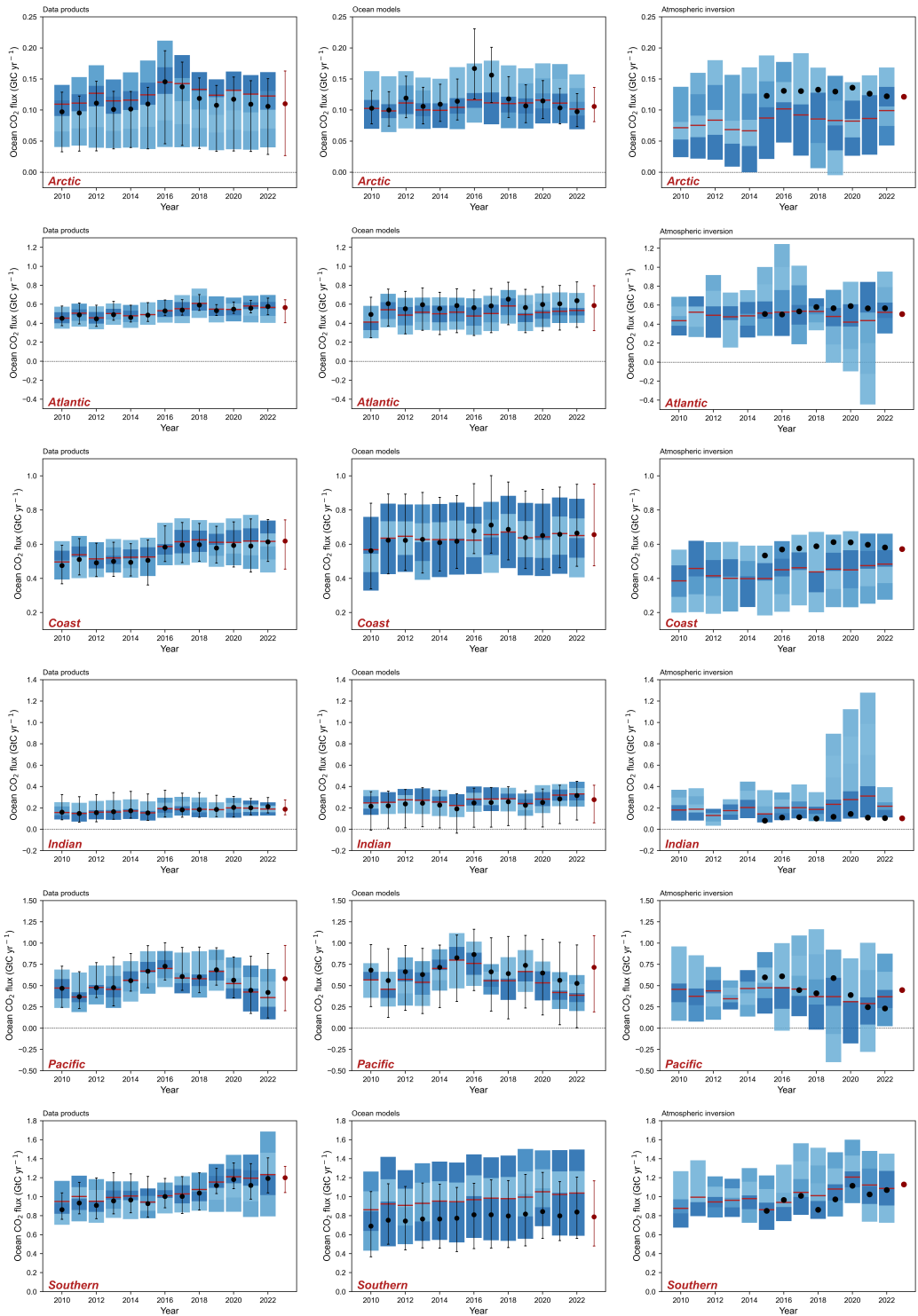

**Supplementary Figure 5 Regional ocean fluxes for the RECCAP2 ocean regions from data products (left column), Global Ocean Biogeochemical Models (middle column) and inversion model (right column).** The blue distribution is from the models used in latest Global Carbon Budget assessments (darker color means more models around a value). The median of models is the red line, the mean of the AI-based emulators or models used in this study is indicated by black dots and by a red dot in 2023. See Supplementary Figure S6 for the distribution of the RECCAP2 regions.

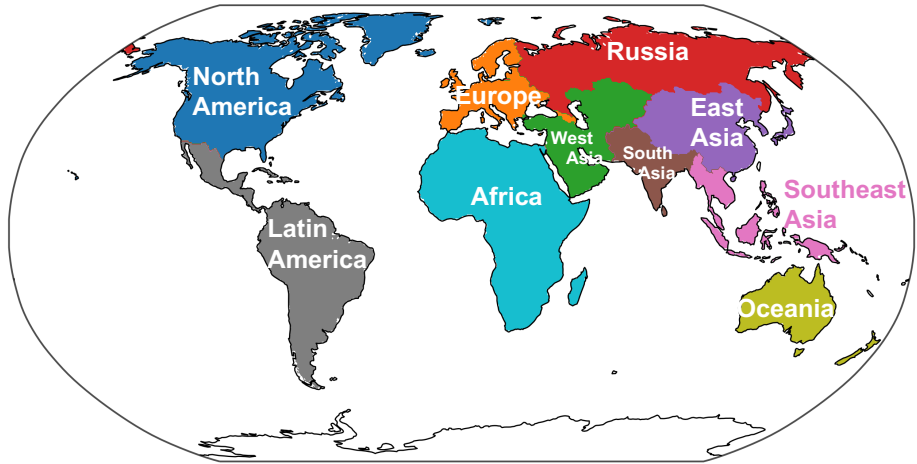

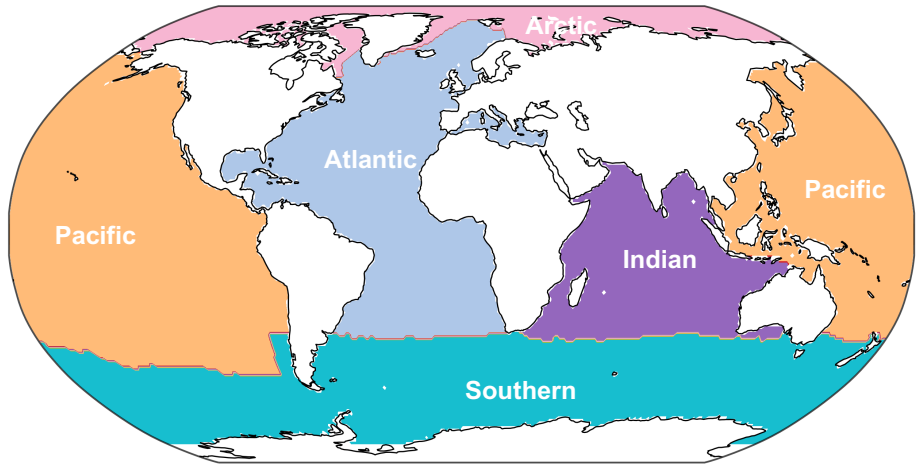

**Supplementary Figure 6** Maps of RECCAP2 land (a) and ocean (b) regions.

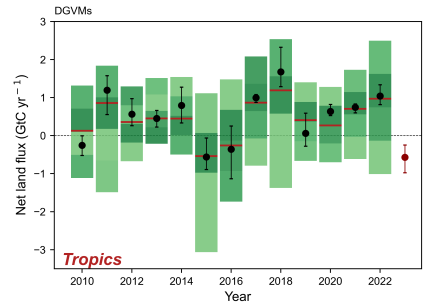
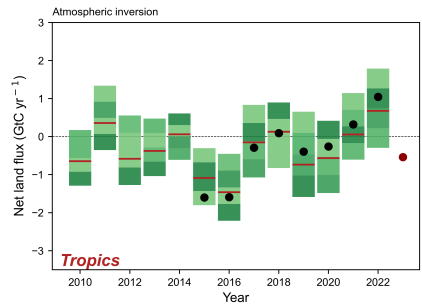
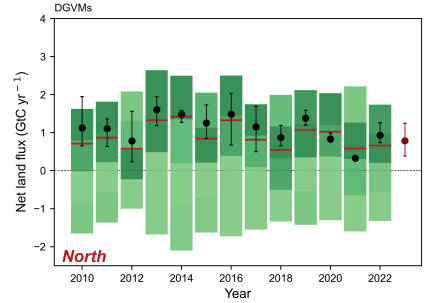
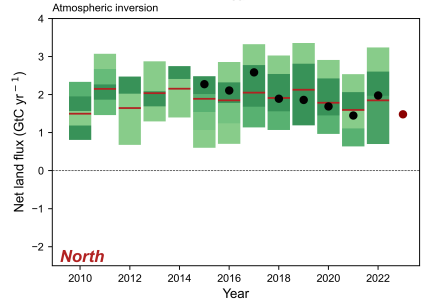
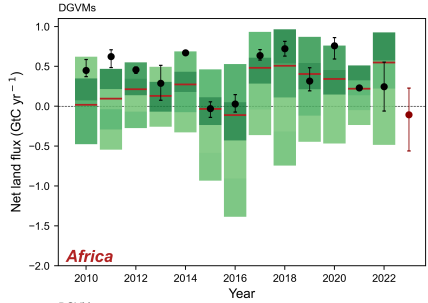
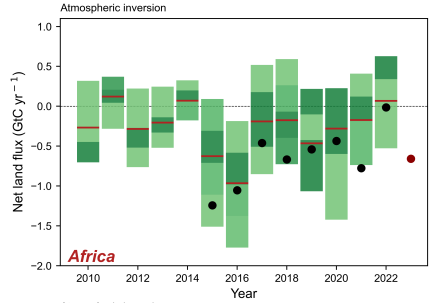
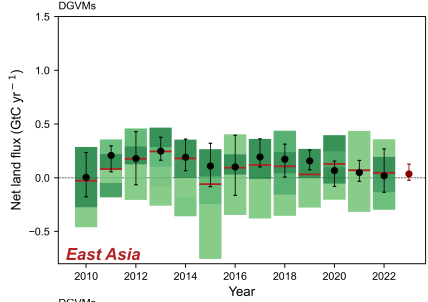
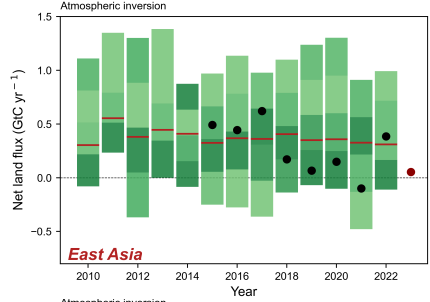
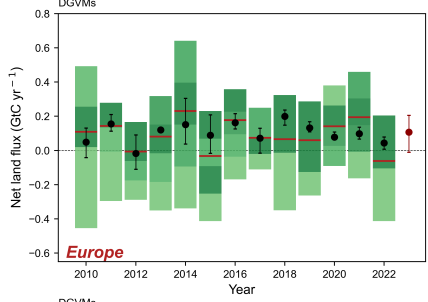
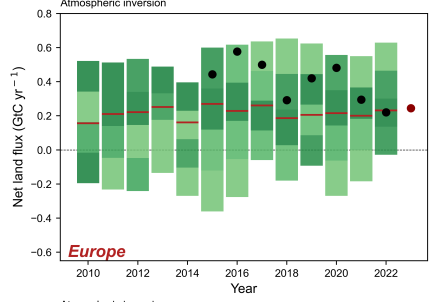
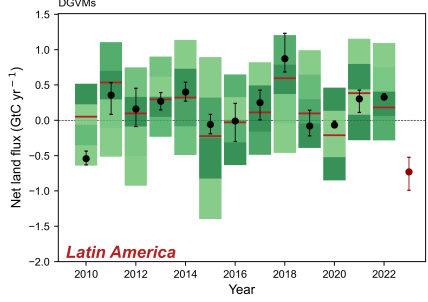
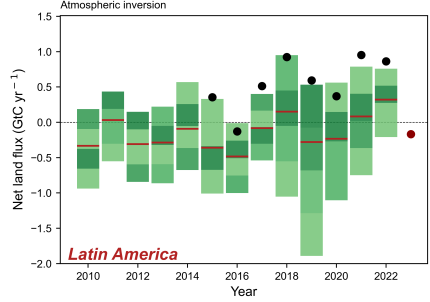

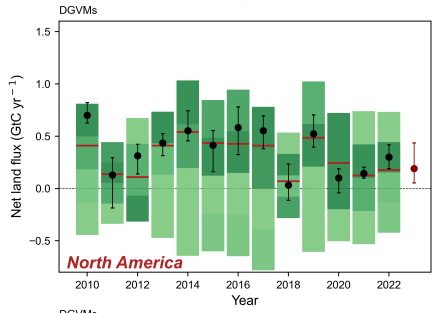
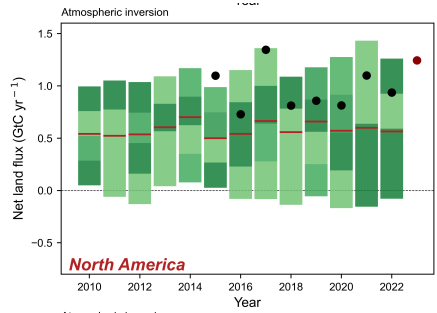
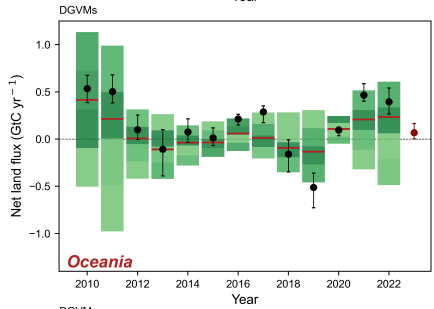
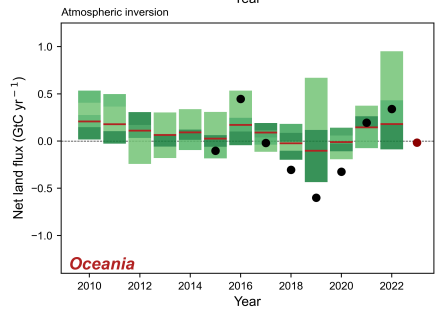
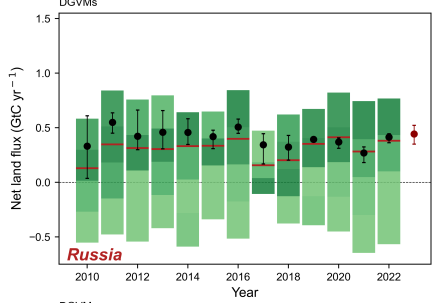
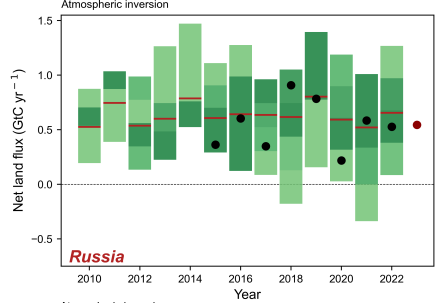
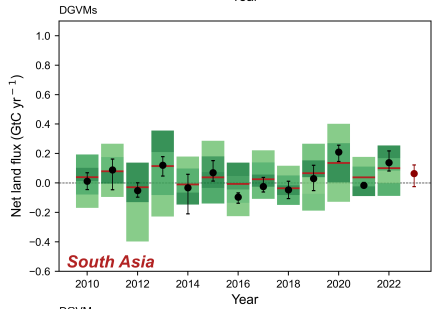
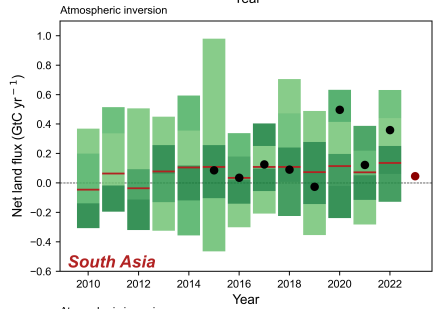
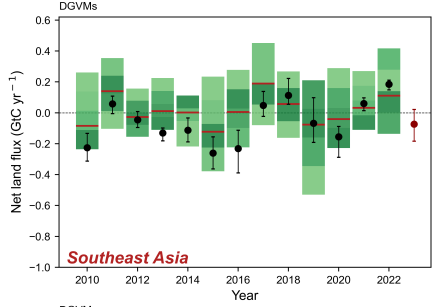
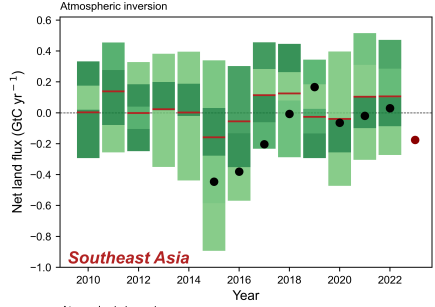
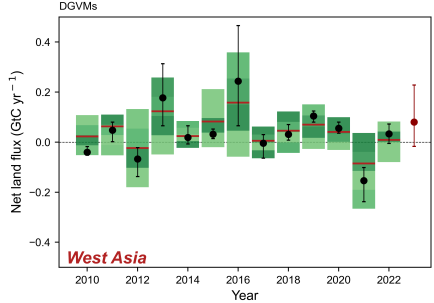
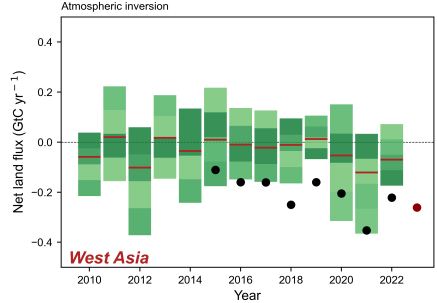

**Supplementary Figure S7 Regional land fluxes for the RECCAP2 land regions from DGVMs (left) and inversion models (right).** The green color distribution is from the TRENDY and inversion models used in latest Global Carbon Budget assessments (darker color means more models around a value). The median of TRENDY and inversion models is the red line, the mean of the 3 DGVMs and OCO-2 inversion used in this study is indicated by black dots and by a red dot in 2023. See Supplementary Figure S6 for the distribution of the RECCAP2 regions.

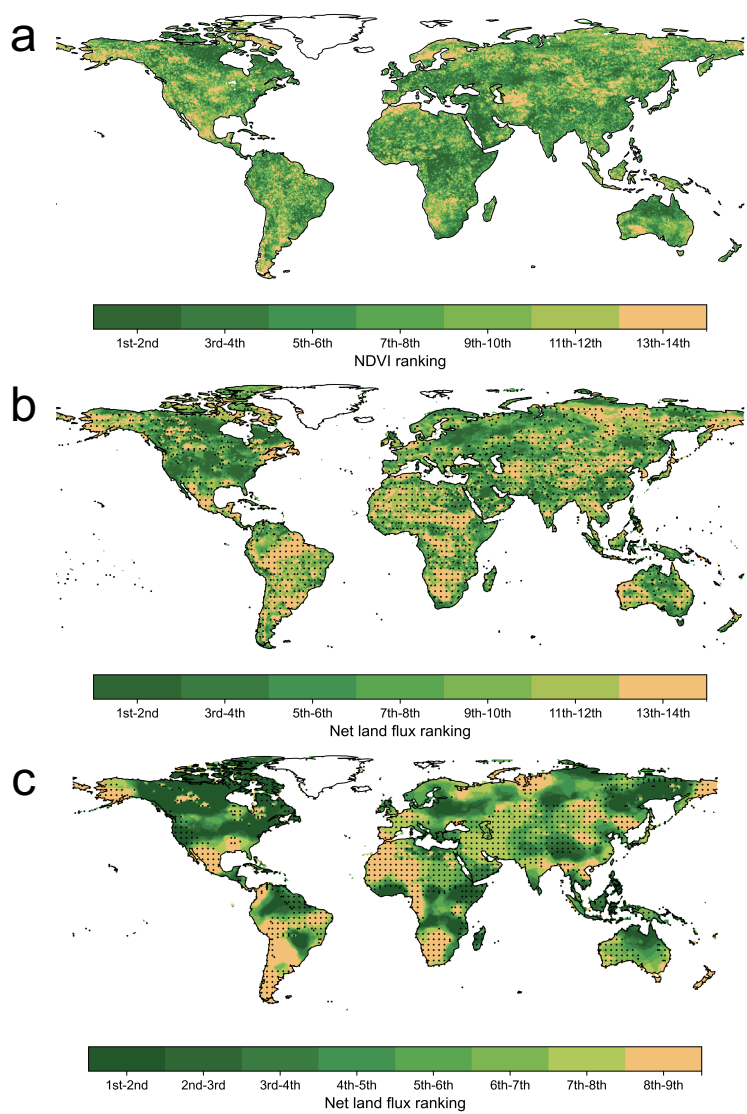

**Supplementary Figure 8 (a) Ranks of the greenness index (NDVI from ref[8]) in each grid cell during 2010-2023 (rank 1 indicates the highest value among the 14 years). (b) Ranks of the bottom-up DGVM fluxes during the same period (rank 1 indicates the highest sink among the 14 years, rank 14 indicates the smallest sink or the largest source; to better appreciate if a pixel of low rank is a source or remains a sink, each pixel that was an annual $CO_2$ source in 2023 is stippled. (c) Ranks of the OCO-2 inversion fluxes during 2015-2023 (rank 1 indicates the highest sink among the 9 years, rank 9 indicates the smallest sink or the largest source.**

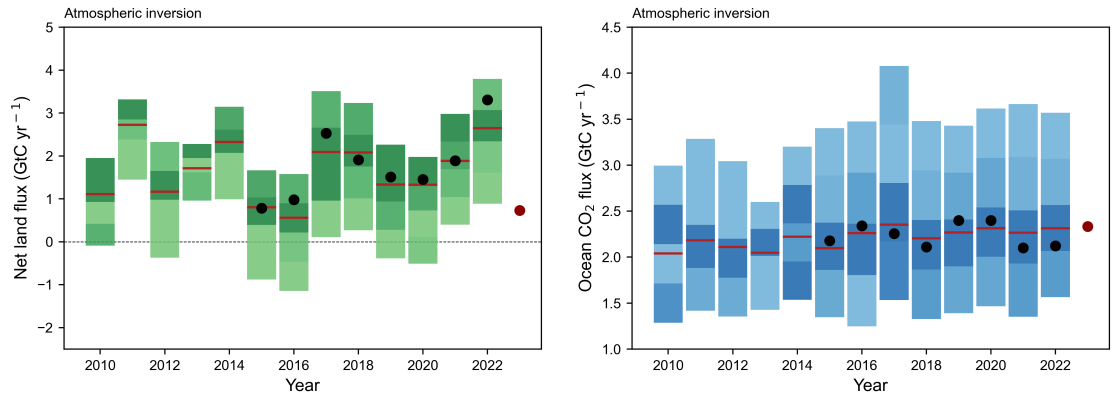

**Supplementary Figure 9** Comparison of global land (left) and ocean (right) CO$_2$ fluxes between the OCO-2 inversion of this study (black dots, red dots for 2023) and the distribution of 14 inversion models (green bars for land, blue bars for ocean, with color intensity representing the number of models falling into each interval) used in ref.[7]

**Supplementary Table 1** $CO_2$ fluxes (2010–2022 average and 2023) for each land RECCAP2 region from DGVMs and inversion methods (2010–2022 from GCB 2023; 2023 from NRT methods; Unit: GtC/yr).

| Period | Methods | World | Tropics | North | Africa | East Asia | Europe | Latin America | North America | Oceania | Russia | South Asia | Southeast Asia | West Asia |
|---|---|---|---|---|---|---|---|---|---|---|---|---|---|---|
| 2010-2022 average | DGVMs | 1.65 ± 2.76 | 0.45 ± 0.94 | 0.90 ± 2.42 | 0.24 ± 0.49 | 0.09 ± 0.36 | 0.09 ± 0.30 | 0.17 ± 0.69 | 0.31 ± 0.81 | 0.06 ± 0.23 | 0.30 ± 0.76 | 0.04 ± 0.18 | 0.01 ± 0.18 | 0.04 ± 0.07 |
| | Inversions | 1.67 ± 1.12 | -0.33 ± 1.13 | 1.89 ± 0.39 | -0.26 ± 0.71 | 0.38 ± 0.18 | 0.24 ± 0.41 | -0.14 ± 0.46 | 0.58 ± 0.12 | 0.09 ± 0.19 | 0.64 ± 0.17 | 0.07 ± 0.12 | 0.03 ± 0.18 | -0.04 ± 0.16 |
| 2023 | DGVMs | 0.14 ± 0.28 | -0.57 ± 0.41 | 0.78 ± 0.46 | -0.11 ± 0.45 | 0.04 ± 0.09 | 0.11 ± 0.12 | -0.73 ± 0.26 | 0.19 ± 0.25 | 0.07 ± 0.10 | 0.44 ± 0.09 | 0.06 ± 0.09 | -0.07 ± 0.11 | 0.08 ± 0.15 |
| | Inversions | 0.73 ± 0.30 | -0.54 ± 0.20 | 1.48 ± 0.15 | -0.66 ± 0.26 | 0.05 ± 0.17 | 0.24 ± 0.07 | -0.17 ± 0.26 | 1.24 ± 0.19 | -0.02 ± 0.10 | 0.54 ± 0.12 | 0.05 ± 0.22 | -0.18 ± 0.22 | -0.26 ± 0.17 |

**Supplementary Table 2** $CO_2$ fluxes (2010–2022 average and 2023) for each ocean RECCAP2 region from bottom-up and inversion methods (2010–2022 from GCB 2023; 2023 from NRT methods; Unit: GtC/yr).

| Period | Methods | World | Arctic | Atlantic | Coast | Indian | Pacific | Southern |
|---|---|---|---|---|---|---|---|---|
| 2010-2022 average | GOBMs | 2.48 ± 0.80 | 0.11 ± 0.06 | 0.51 ± 0.18 | 0.64 ± 0.22 | 0.27 ± 0.12 | 0.58 ± 0.24 | 0.97 ± 0.43 |
| | Data products | 2.46 ± 0.66 | 0.13 ± 0.09 | 0.52 ± 0.14 | 0.57 ± 0.13 | 0.18 ± 0.09 | 0.53 ± 0.27 | 1.06 ± 0.27 |
| | Inversions | 2.21 ± 0.17 | 0.08 ± 0.02 | 0.49 ± 0.07 | 0.44 ± 0.05 | 0.20 ± 0.11 | 0.40 ± 0.11 | 1.01 ± 0.20 |
| 2023 | GOBMs | 2.49 ± 0.79 | 0.11 ± 0.03 | 0.59 ± 0.26 | 0.66 ± 0.30 | 0.28 ± 0.22 | 0.71 ± 0.52 | 0.79 ± 0.38 |
| | Data products | 2.67 ± 0.60 | 0.11 ± 0.08 | 0.57 ± 0.16 | 0.62 ± 0.16 | 0.18 ± 0.09 | 0.58 ± 0.39 | 1.20 ± 0.16 |
| | Inversions | 2.33 ± 0.20 | 0.12 ± 0.20 | 0.51 ± 0.20 | 0.57 ± 0.20 | 0.10 ± 0.20 | 0.45 ± 0.20 | 1.13 ± 0.20 |

**Supplementary Table 3** Quarterly differences of four versions of OCO-2 inversions in 2023 compared to quarterly means from 2015 to 2022 (Unit: GtC/yr).

| Region | Quarter | Inversion without fires | Inversion with fires but regrowth | Inversion with fires without regrowth |
|---|---|---|---|---|
| World | Q1 | 0.066 | 0.070 | 0.030 |
| | Q2 | 0.070 | 0.050 | 0.115 |
| | Q3 | -0.229 | -0.243 | -0.248 |
| | Q4 | -0.290 | -0.243 | -0.252 |
| North America > 60°N | Q1 | 0.000 | 0.009 | 0.000 |
| | Q2 | 0.018 | 0.022 | 0.019 |
| | Q3 | 0.005 | -0.009 | 0.007 |
| | Q4 | 0.000 | 0.011 | 0.003 |
| Europe > 60°N | Q1 | 0.001 | 0.000 | 0.002 |
| | Q2 | 0.012 | 0.014 | 0.007 |
| | Q3 | -0.009 | -0.007 | -0.008 |
| | Q4 | -0.004 | -0.004 | -0.004 |
| Siberia > 60°N | Q1 | -0.001 | -0.006 | 0.000 |
| | Q2 | -0.032 | -0.034 | -0.002 |
| | Q3 | 0.038 | 0.051 | 0.041 |
| | Q4 | 0.001 | -0.003 | -0.004 |